\documentclass[graybox]{svmult}

\usepackage{mathptmx}       % selects Times Roman as basic font
\usepackage{helvet}         % selects Helvetica as sans-serif font
\usepackage{courier}        % selects Courier as typewriter font
\usepackage{type1cm}        % activate if the above 3 fonts are
\usepackage{makeidx}         % allows index generation
\usepackage{graphicx}        % standard LaTeX graphics tool
\usepackage{multicol}        % used for the two-column index
\usepackage[bottom]{footmisc}% places footnotes at page bottom

\makeindex             % used for the subject index
\begin{document}

\title*{Lopsidedness in WHISP galaxies}
\author{E. J\"utte, J. van Eymeren, Ch. Jog, R.-J. Dettmar, Y. Stein}
\institute{Astronomisches Institut der Ruhr--Universit\"at Bochum (AIRUB), \email{eva.juette@astro.rub.de}
\and Fakult\"at f\"ur Physik, Universit\"at Duisburg-Essen  \and Department of Physics, Indian Institute of Science \and AIRUB \and AIRUB}
\maketitle

% Too much empty space in the original style file!
\vskip-1.2truein

\abstract{Observations of the stellar and gaseous components in disc galaxies often reveal asymmetries in the morphological and kinematic distribution. However, the origin of this effect is not well known to date, and quantitative studies are rare. Here, we present the first statistical investigation of a sample of 76 H{\sc i} discs using the WHISP survey. We perform a Fourier analysis to study the morphological lopsidedness. This allows to trace the degree of asymmetry with radius. We further investigate the dependence on, e.g., the morphological type and the environment. 
}

\section{Introduction}
\label{sec:1}
It has been known for a long time that there are non-axisymmetric characteristics in discs of spiral galaxies in the stellar and gaseous morphology and/or kinematics and therefore in the mass distribution (e.g., \cite{baldwin80,rix95,jog09}). Lopsided galaxies show a global asymmetry so that the mass distribution can be characterised by the Fourier amplitude $m=1$. Lopsidedness is particularly seen in the outer optical discs of 50\% of the galaxies (e.g., \cite{rix95,richter94}). Most of the studies, however, are based on optical imaging or global H{\sc i} profiles. H{\sc i} maps are ideal to investigate lopsidedness since gas discs are much further extended than stellar discs. Also, both morphological and kinematic information are provided at the same time. 
 
Possible scenarios for lopsidedness are tidal interaction \cite{jog97} or minor merger \cite{zaritsky97}, continuous gas accretion \cite{bournaud05}, but also an offset of the stellar disc in a halo potential \cite{noordermeer01}. 

We used a sub-sample of the WHISP\footnote{Westerbork H{\sc i} Survey of Irregular and Spiral galaxies} survey to systematically analyse a large set of H{\sc i} discs with respect to the local and global morphology. Our sample includes 76 galaxies which were chosen according to the following selection criteria: (1) the ratio of the H{\sc i} diameter over  the beam size has to be larger than 10 at a resolution of 30 arcsec; (2) the galaxies need to have inclinations between 20$\,^{\circ}$ and 75$\,^{\circ}$. This sample covers a whole range of morphological types and galaxy masses. We trace the neutral gas out to at least twice the optical radius, sometimes even further.

\section{Method \& Results}
\label{sec:2}
Lopsidedness can be described as $m=1$ mode of a Fourier decomposition \cite{rix95}:
\begin{equation}
\sigma(r,\phi) = a_0 (r) + \sum a_m (r) \cdot cos [m\cdot \phi' - \phi_m (r)]
\end{equation}
Here, $a_0$ is the mean surface density, $a_m$ the amplitude of the surface density harmonic coefficient, $\phi_m$ the phase and $\phi'$ the azimuthal angle in the plane of the galaxy. Lopsidedness can then be characterised by the normalised Fourier amplitude $A_1 = a_1/a_0$ and the phase $\phi_1$. We performed a harmonic decomposition of the H{\sc i} intensity maps using the kinematic parameters derived from a tilted-ring analysis.

\begin{figure}[b]
%
%\sidecaption
% For example, with the graphicx style use
\includegraphics[scale=.70]{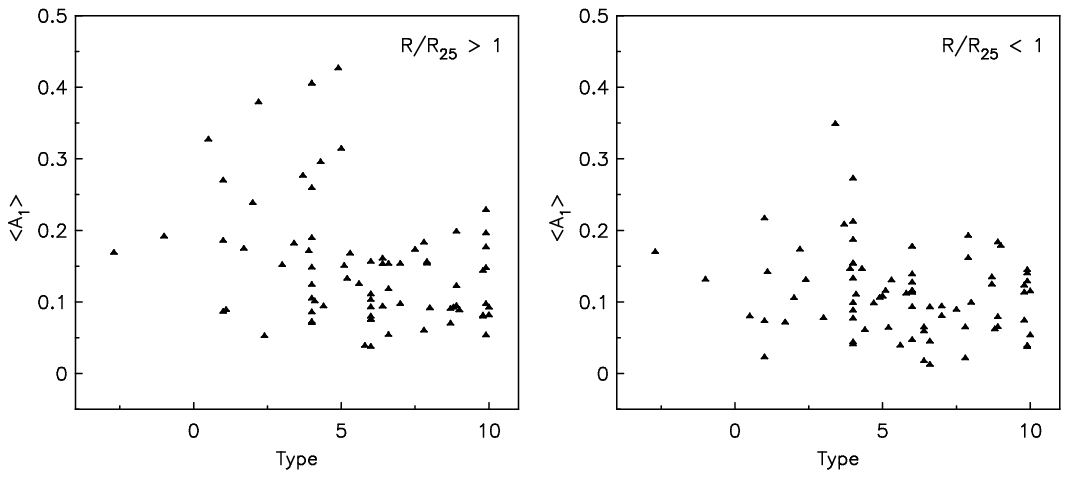}
\includegraphics[scale=.70]{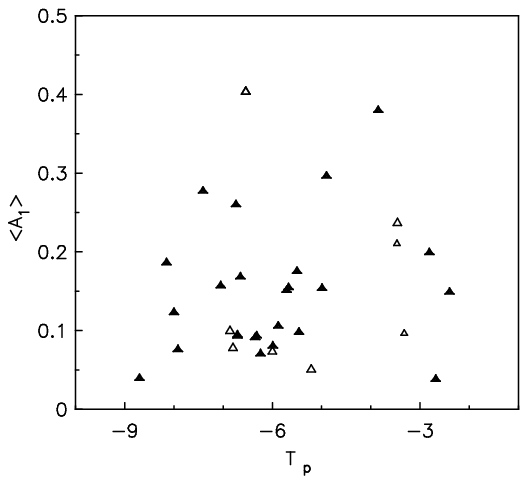}
\caption{(left) $<A_1>$ \emph{vs.} morphological type averaged over large and small radii. (right) $<A_1>$ \emph{vs.} tidal parameter.}
\label{fig:1}       % Give a unique label!
\end{figure}

We found that within the optical disc $A_1$ increases continuously with radius, but seems to saturate beyond that. Early-type disc galaxies tend to show a higher lopsidedness than late-type galaxies, in particular in the outer parts (Fig.~\ref{fig:1}, left panels). This is in agreement with \cite{angiras06}. We calculated the tidal parameter in order to investigate the correlation of lopsidedness with the environment  \cite{bournaud05}. As Fig.~\ref{fig:1}, right panel shows, lopsidedness and the environment of the sample galaxies are not correlated (in agreement with \cite{bournaud05}). Both results together indicate a tidal origin for the lopsidedness: the disc responds to a halo that is distorted by a tidal encounter \cite{jog97}. For a detailed analysis see \cite{eymeren11}.
%We found that 20\% of our sample show a continuous increase of $A_1$ with radius, which we consider as being lopsided.\\
%Early-type disc galaxies tend to show a higher lopsidedness, in particular in the outer parts (\ref{fig:1}.
%This is in agreement with Angiras et al. (2006). \\
%Furthermore, we investigated the correlation of lopsidedness and the environment. The environment was characterised by the tidal parameter as defined in Bournaud et al. 2005. We do not find a correlation between lopsidedness and the environment. Thus, we conclude that tidal interaction is not the dominating origin of lopsidedness.

%%% Bibliography

%\input{referenc}
\end{document}